\documentclass{emulateapj}
\usepackage{graphicx}

\slugcomment{ }

\begin{document}
\title{Kilonova emission from black hole-neutron star mergers: observational signatures of anisotropic mass ejection}

\author{Zhi-Qiu Huang\altaffilmark{1,2}, Liang-Duan Liu\altaffilmark{1,2,3}, Xiang-Yu Wang\altaffilmark{1,2}, Zi-Gao Dai\altaffilmark{1,2}}

\altaffiltext{1}{School of Astronomy and Space Science, Nanjing University, Nanjing 210093, China;
xywang@nju.edu.cn,dzg@nju.edu.cn}
\altaffiltext{2}{Key laboratory of Modern Astronomy and Astrophysics (Nanjing University), Ministry of Education, Nanjing 210093, China}
\altaffiltext{3}{Department of Physics and Astronomy, University of Nevada,Las Vegas,NV 89154,USA}
\begin{abstract}
The gravitational wave event GW170817 associated with the short gamma-ray burst (GRB) 170817A confirms that  binary neutron star (BNS) mergers are one of the origins of short GRBs. The associated kilonova emission, radioactively powered by nucleosynthesized heavy elements, was also detected. Black hole-neutron star (BH-NS) mergers have been argued to be another promising origin candidate of short GRBs and kilonovae. Numerical simulations show that the ejecta in BH-NS mergers is geometrically much more anisotropic than the BNS merger case. In this paper, we investigate observational signatures of kilonova emission from the anisotropic ejecta in BH-NS mergers. We find that a bump appears on the bolometric luminosity light curve due to the inhomogeneous mass distribution in the latitudinal direction. The decay slope of the single-band light curve becomes flatter and the spectrum also deviates from a single-temperature blackbody radiation spectrum due to the gradient in the velocity distribution of the ejecta. Future detection or non-detection of such signatures would be useful to test the mass ejection geometry in BH-NS mergers.

\end{abstract}


\section{Introduction}
 The ejecta from binary neutron star (BNS) mergers or black hole-neutron star (BH-NS) mergers has long been suggested to be the source for rapid neutron capture (\emph{r}-process) nucleosynthesis producing elements heavier than iron (e.g. Lattimer \& Schramm 1974). \cite{1998ApJ...507L..59L} proposed that the radioactive ejecta of a BNS merger could power a
supernova-like thermal transient. \cite{2010MNRAS.406.2650M} calculated  the heating rate of the ejecta  based on numerical simulations through a detailed nuclear reaction network. They invoked a physical model for the opacity and  predicted a peak luminosity  of  roughly 1000 times more
luminous than classical novae, so that the events were dubbed as 'kilonova'. Later, more detailed studies on the characteristics of the kilonova emission were performed \citep[e.g.,][]{2011ApJ...736L..21R,2013ApJ...774...25K,
 2013ApJ...775...18B,2013ApJ...775..113T,2014MNRAS.441.3444M,2014ApJ...789L..39W,2014MNRAS.443.3134P,2015MNRAS.450.1777K,
 2015ApJ...813....2M,2016ApJ...819..120L,2017arXiv171005931M}.

 The gravitational wave event from a binary neutron star merger, known as GW170817, was detected on August 17, 2017 at 12:41:04 UTC, by LIGO/Virgo. A short gamma-ray burst (SGRB 170817A) was observed by the \emph{Fermi} Gamma-Ray Telescope and the International Gamma-Ray Astrophysics Laboratory (INTEGRAL) $1.74\,$s later post-merger \citep{2017ApJ...848L..13A,2017ApJ...848L..14G,2017ApJ...848L..15S,2018NatCo...9..447Z}. A kilonova, namely AT 2017gfo, was also detected $\sim 10$\,hours after this merger \citep[e.g.,][]{2017Natur.551...64A,2017Sci...358.1556C,2017Sci...358.1570D,2017Sci...358.1565E,
 2017ApJ...848L..19C,2017NatAs...1..791C,2017ApJ...848L..17C,2017SciBu..62.1433H,2017Sci...358.1559K,
 2017Sci...358.1583K,2017ApJ...848L..18N,2017Natur.551...67P,2017Sci...358.1574S,2017Natur.551...75S,
 2017ApJ...848L..16S,2017ApJ...848L..27T,2017PASA...34...69A}.  The kilonova emission is considered to be characterized by components with different opacities, including a blue kilonova that decays rapidly, a purple kilonova with an intermediate opacity, and a red kilonova that evolves slowly due to lanthanide-rich materials in ejecta \citep{2014MNRAS.441.3444M,2015MNRAS.450.1777K,2017ApJ...848L..17C,2017Sci...358.1570D,2017Natur.551...80K,2017arXiv171005931M,2017ApJ...850L..37P,
 2017ApJ...851L..21V}.

Most calculations on the kilonova emission from BNS mergers assumed an almost isotropic mass distribution of the ejected mass in the longitudinal and the latitudinal directions\citep[e.g.,][]{2016ApJ...825...52K,2017ApJ...850L..41X,2018ApJ...861..114Y}, which may be reasonable for BNS mergers. The observed low degree of polarization of the kilonova emission in GW170817 suggests a nearly-symmetric geometry of the ejecta \citep{2017NatAs...1..791C}, which is consistent with some numerical simulations on dynamical ejecta from BNS mergers\footnote{\cite{2017ApJ...850L..37P} considered an anisotropy geometry for AT 2017gfo, but the degree of  the anisotropy is still much lower than the case of BH-NS mergers.} \citep[e.g.,][]{2017CQGra..34j5014D}.
However, the situation is drastically different for BH-NS mergers. Numerical simulations have shown that the geometry of the dynamic ejecta from BH-NS mergers is highly anisotropic \citep[e.g.,][]{2013PhRvD..88d1503K,2015PhRvD..92d4028K,2016ApJ...825...52K}. Geometrically, the dynamic ejecta exhibits a crescent-like shape, i.e., the ejecta is concentrated around the orbital plane with a small polar opening angle and sweeps out half of the plane.
\cite{2015PhRvD..92d4028K} and \cite{2016ApJ...825...52K}  calculate the kilonova emission from BH-NS mergers by  considering this anisotropic geometry. These works have assumed a homogeneous density distribution in the latitudinal direction.  Differently from these earlier works, we here consider the inhomogeneous density distribution in the latitudinal direction, as found in some simulations \citep[e.g.,][]{2015PhRvD..92d4028K}. In addition, the velocity of the ejecta may span a wide range, which in turn affects the density structure of the ejecta. We  consider a power-law velocity distribution of the ejecta,  motivated by recent studies on the kilonova emission of GW170817A. We find that interesting signatures appear in the kilonova emission when these effects are take into account.

This paper is organized as follows. In Section 2, we describe the model for calculating the kilonova emission when   the anisotropy in geometry is taken into account. In Section 3, our results are shown and compared with the previous models. We use our model to constrain the mass ejecta of GRB130603B in Section 4. In Section 5, we present our discussions and conclusions.

\section{Model}


\begin{figure*}
\begin{center}
\includegraphics[scale=0.34]{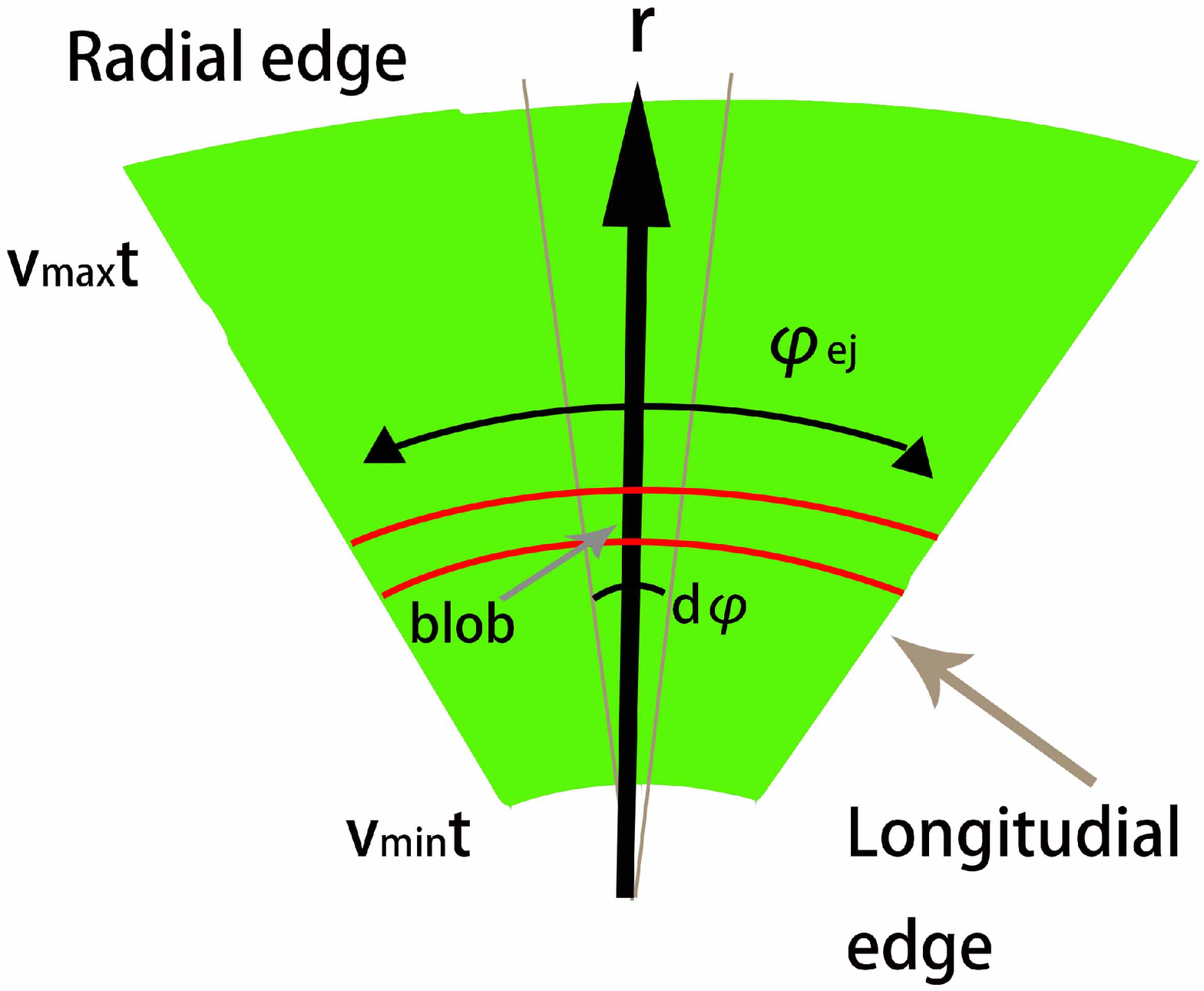}
\includegraphics[scale=0.36]{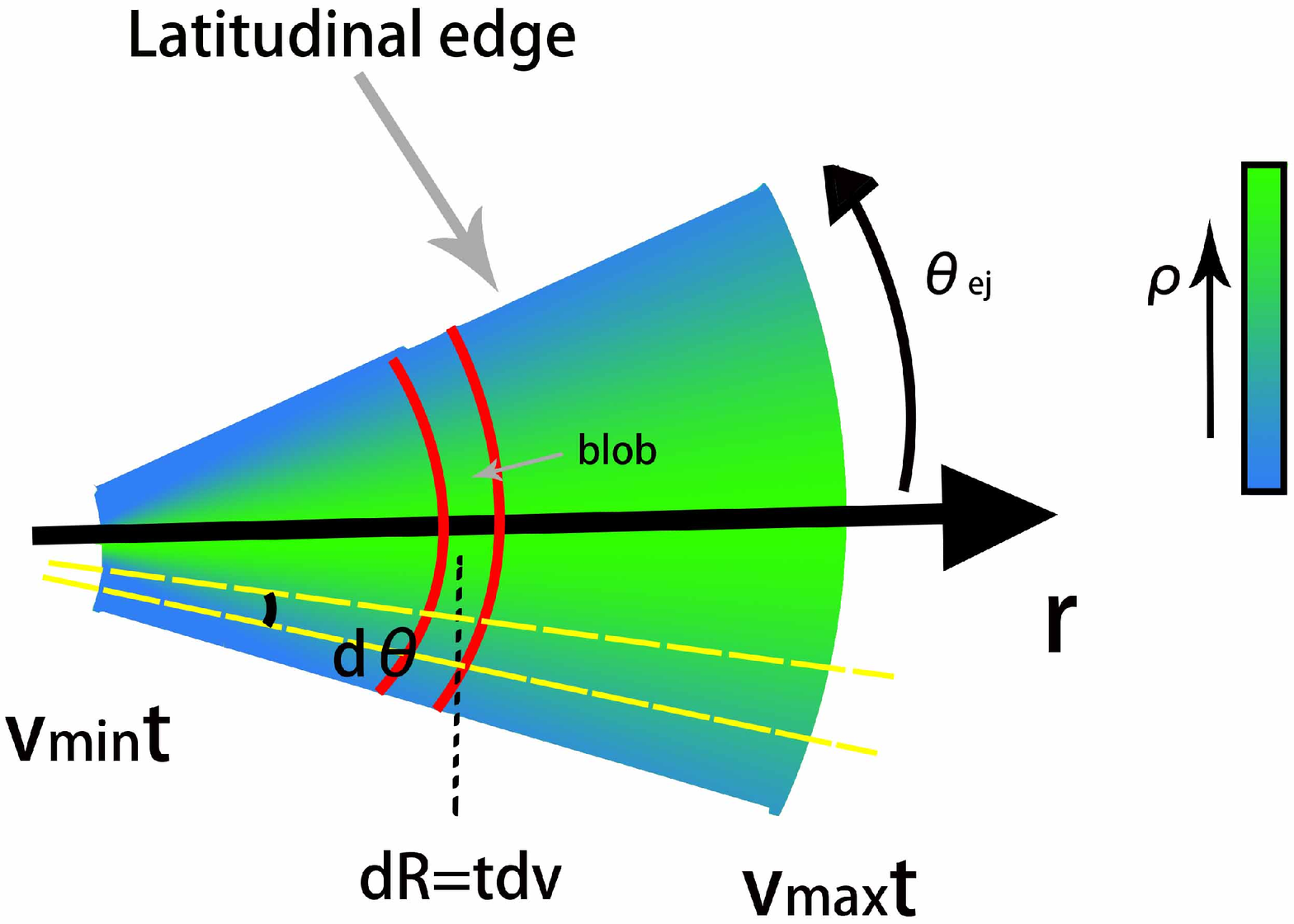}
\caption{A schematic picture of the ejecta distribution profile. $v_{\rm{max}}$ and $v_{\rm{min}}$ are the highest and lowest velocity. The left panel shows
the structure in the longitudinal direction and $\varphi_{\rm{ej}}$ is the azimuthal opening angle. The right panel shows the structure in the latitudinal direction and $\theta_{\rm{ej}}$ is the half thickness of the ejecta. The density is homogenous in the longitudinal direction while it decreases in the latitudinal direction with $\theta$. The original picture is shown as Fig. 3 in \cite{2016ApJ...825...52K}. The latitudinal structure is modified here to show the density gradient along the latitudinal direction, which is considered in our calculation.}
\end{center}
\end{figure*}

The geometry of the dynamical ejecta of BH-NS mergers can be described by a partial sphere in the latitudinal and longitudinal directions \citep{2013PhRvD..88d1503K,2015PhRvD..92d4028K,2016ApJ...825...52K,2017ApJ...850L..41X}, as shown in Fig. 1. According to the simulations by \cite{2015PhRvD..92d4028K}, the mass density in the latitudinal direction decreases with the latitudinal angle, so we assume that the mass distribution in the latitudinal direction follows
\begin{equation}
\frac{dm}{\cos \theta d\theta}\propto
\left\{
\begin{array}{ll}
1, & {\rm if}\,\,\theta\leq \theta_{c}, \\
(\frac{\theta}{\theta_{c}})^{k}, & {\rm if}\,\,\theta>\theta_{c},%
\end{array}%
\right.
\end{equation}
where $\theta_{c}=0.05$ is assumed according to the numerical simulations in \cite{2015PhRvD..92d4028K}. $dm$ is the mass of   the material with a certain velocity in a certain longitudinal direction. The normalization is calculated by $\int_{0}^{\theta_{\rm{ej}}} 2\frac{dm}{\cos \theta d\theta}d\theta=\frac{dM}{dv}/\varphi_{\rm{ej}}$.

After the merger, the ejecta velocity structure approaches the homologous expansion, with the faster matter lying ahead of the slower matter (e.g.,
Bauswein et al. 2013; Rosswog et al. 2014). A power-law evolution of the photospheric radius in GW170817/GRB170817A, $r_{\rm{ph}} \propto t^{0.6}$, suggests a power- law mass distribution in velocity \citep[e.g.,][]{2017arXiv171109638W}. Motivated by this,  we assume the mass distribution of the dynamic ejecta in BH-NS mergers also follows a power-law distribution, i.e.,
\begin{equation}
\frac{dM}{dv}\propto v^{s},
\end{equation}
with $s\le 0$. The normalization is determined by $\int_{v_{\rm{min}}}^{v_{\rm{max}}}\frac{dM}{dv}dv=M_{\rm{ej}}$. Numerical simulations have shown the bulk velocity of the ejecta is $v_{\rm{ej}} \equiv P_{\rm{ej}}/M_{\rm{ej}} \sim 0.2c$, where $P_{\rm{ej}}$ is the bulk momentum \citep{2013PhRvD..88d1503K, 2015PhRvD..92d4028K}. $P_{\rm{ej}}$ can be obtained by
\begin{equation}
P_{\rm{ej}} = \int _{v_{\rm{min}}}^{v_{\rm{max}}} v \frac{dM}{dv} dv.
\end{equation}
For a homogeneous mass distribution in velocity, i.e., $\frac{dM}{dv} = {\rm{constant}}$, we  obtain $v_{\rm{min}}=0.1c$ for a maximum velocity  of $v_{\rm{max}}=0.3c$, which are consistent with the values adopted in \citet{2017LRR....20....3M}. {{It is worthy noted that  the minimum velocity of ejecta $v_{\rm min}$ depends on the value of $s$. For $s=-1$, then one  obtains $v_{\rm min}=0.13c$.}}

We consider that the ejecta is composed by blobs with different velocities in a certain longitudinal and latitudinal directions. The volume of the blob is given by
\begin{equation}
dV=r \cos \theta d\theta \cdot rd\varphi \cdot tdv,
\end{equation}
where we assume a homologous expansion for the blob  and $r=vt$.
The normalization is given by
\begin{equation}
\int_{0}^{\varphi_{\rm{ej}}} \int_{v_{\rm{min}}}^{v_{\rm{max}}} \int_{0}^{\theta_{\rm{ej}}} 2\rho \cos \theta v^{2} t^{3} d\varphi dv d\theta = M_{\rm{ej}}.
\end{equation}
Following the simulations by \cite{2015PhRvD..92d4028K}, we take $\varphi_{\rm{ej}}=\pi$ and $\theta_{\rm{ej}}=0.2$.
It is easily found that
\begin{equation}
\varphi_{\rm{ej}} \int_{0}^{\theta_{\rm{ej}}} 2\rho \cos \theta v^{2} t^{3} d\theta=\frac{dM}{dv}.
\end{equation}

Note that $\varphi _{\rm{ej}}$ and $\theta _{\rm{ej}}$ may depend on the equation of state, the mass ratio between NS and BH, and the spin. However, \citet{2015PhRvD..92d4028K} found the dependence on these factors are weak and the expected uncertainties are less than 25 percent. Thus, we neglect these factors.

Then we can obtain the density of the blob,
\begin{equation}
\rho=\frac{dm}{\cos \theta d\theta(v^{2}t^{3})}.
\end{equation}

Considering the random walk of photons \citep{2016ApJ...825...52K}, the depth of the visible mass is determined by the condition that the distance to the latitudinal edge is comparable to the distance that a photon diffuses,
namely $vt \left( \theta_{\rm{ej}}-\theta_{\rm{obs}} \right)\approx ct/\tau$. Here, $\tau\approx \int_{\theta_{\rm{obs}}}^{\theta_{\rm{ej}}} \kappa \rho \cos \theta vtd\theta$, and $\theta_{\rm{obs}}$ is the depth of the visible mass at a certain time. When $\theta_{\rm{obs}}=0$, the whole part of the ejecta can be seen. The opacity of r-process ejecta is much higher than that for Fe group elements \citep{2013ApJ...774...25K,2013ApJ...775...18B}, with $\kappa \sim 10-100 \, \rm{cm^{2} g^{-1}}$. Here, $\kappa=10\,\rm{cm^{2} g^{-1}}$ is adopted as a reference value.

For a blob with a certain velocity in a certain longitudinal direction, the visible mass is
\begin{equation}
\frac{dM_{\rm{obs}}}{d\varphi dv}=2\int_{\theta_{\rm{obs}}}^{\theta_{\rm{ej}}} \rho r^2 t \cos \theta d\theta.
\end{equation}

We assume that the emission from the kilonova is produced by radioactive decay without any additional energetic engine \citep{2013MNRAS.430.2121P,2016ApJ...825...52K,2017ApJ...850L..41X}. The radioactive heating rate is approximated by a power law $\epsilon(t)=\epsilon_{0} (t/{\rm day})^{-\alpha}$ \citep{2012MNRAS.426.1940K,2014ApJ...789L..39W}. The bolometric luminosity of this blob is
\begin{equation}
\frac{dL}{d\varphi dv}=(1+\theta_{\rm{ej}})\epsilon_{\rm{th}} \epsilon_{0} \frac{dM_{\rm{obs}}}{d\varphi dv} \left(\frac{t} {\rm{day}}
\right)^{-\alpha},
\end{equation}
where $\epsilon_{\rm{th}}=0.5$ is the efficiency of thermalization \citep{2010MNRAS.406.2650M}, $\epsilon_{0}=1.58\times10^{10} \rm{ergs^{-1}g^{-1}}$, and $\alpha=1.3$ \citep{2017CQGra..34j5014D}. The factor $(1+\theta_{\rm{ej}})$ is introduced to include the contribution from the radial edge \citep{2016ApJ...825...52K,2017ApJ...850L..41X}.

Thus, the total bolometric luminosity is
\begin{equation}
L=\int_{0}^{\varphi_{\rm{ej}}} \int_{v_{\rm{min}}}^{v_{\rm{max}}} \frac{dL}{d\varphi dv} d\varphi dv.
\end{equation}
Assuming the spectrum of the blob as a blackbody,  the temperature of this blob is
\begin{equation}
T_{\rm{eff}}=\left(\frac{dL}{\sigma_{\rm{SB}} 2vt^2 dv d\varphi}\right)^{1/4},
\end{equation}
where $\sigma_{\rm{SB}}$ is the Stephan-Boltzmann constant, and $vt^{2}dv d\varphi $ is the surface area of the blob.

Here we study whether the radiation transfer between adjacent blobs is important or not, as radiation transfer may affect the temperature distribution of the ejecta. 
The radiation transfer flux can be described as
\begin{eqnarray}
F_{rad}&& =  - \frac{4ac}{3\rho} T^3 \frac{dT}{dr} \frac{1}{\kappa}\\ \nonumber
&& \sim - \frac{4ac}{3\rho} T^3 \frac{\Delta T}{\Delta r} \frac{1}{\kappa},
\end{eqnarray}
where $a=4\pi \sigma _{\rm{SB}} /c$ and $\Delta r= t\cdot \Delta v$.

The timescale that the total thermal energy is carried out by radiation in a blob can be estimated by
\begin{equation}
t_{rad}= aT^4 V/(|F_{rad}| S),
\end{equation}
where $V\sim v^2 t^3 \theta _{\rm{ej}} \varphi _{\rm{ej}} \Delta v$ is the volume of the blackbody blob and $S\sim v^2 t^2 \theta _{\rm{ej}} \varphi _{\rm{ej}}$ is the area of the contact surface between blobs.

As an example, considering a blob with $\Delta v= 0.01c$ ranging from $0.1c$ to $0.11c$,  $t_{rad}$ is $\sim 200 {\rm{d}}$ at early time, much longer than the dynamic timescale (i.e., 25 days). Therefore, the effects of radiation transfer can be ignored and each blob can be treated as an adiabatic system, which is consistent with the treatment in \citet{2013ApJ...775..113T}. 

The observed flux at frequency $\nu$ contributed by this  blob is
\begin{equation}
\frac{dF_{\nu}}{d\varphi dv }=\frac{8\pi h \nu^{3}}{c^{2}} \frac{1}{exp(\frac{h\nu}{k_{\rm{B}}T_{\rm{eff}}})-1} \frac{vt^{2}}{4\pi D_{L}^{2}},
\end{equation}
where $D_{L}$ is the luminosity distance. The total flux is then given by
\begin{equation}
F_{\nu}=\int_{0}^{\varphi_{\rm{ej}}} \int_{v_{\rm{min}}}^{v_{\rm{max}}} \frac{dF_{\nu}}{d\varphi dv }   d\varphi dv.
\end{equation}

\section{Results}
Fig. 2 shows the light curves of the bolometric luminosity for various parameters. It can be seen that the mass distribution in the latitudinal direction (the parameter $k$) plays an important role in shaping the bolometric light curves at early times. After all of the ejecta becomes transparent, the light curves merge to the same line. The most interesting result is a clear bump on the lightcurve if the mass distribution in the latitudinal direction is anisotropic ($k \neq 0$). The more anisotropic the mass distribution is, the more remarkable the bump is.

\begin{figure}
\begin{center}
\includegraphics[scale=0.4]{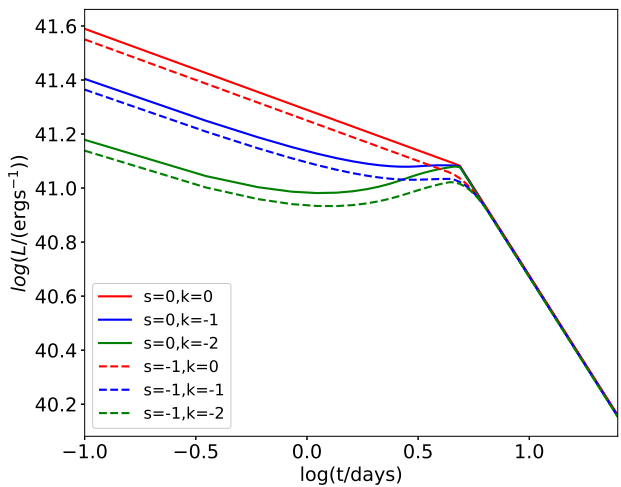}
\caption{Light curves of the bolometric luminosity of the kilonova  with various parameters for the mass distribution in the latitudinal direction and the velocity distribution. $s$ is the power-law index of the mass distribution with velocity, and $k$ is the power-law index of the mass distribution with  the latitudinal angle.}
\end{center}
\end{figure}

The bolometric luminosity mainly depends on the increase of $M_{\rm{obs}}$ with time, which is shown in Fig 3. $M_{\rm{obs}}$ increases more rapidly with time in the case with a smaller $k$ due to the smaller matter density near the latitudinal edge. The fast increase of $M_{\rm{obs}}$ causes the bump on the light curves.

The time when all of the mass of a blob with a certain velocity in a certain longitudinal direction becomes transparent, $t_{\rm{c}}$, is independent of the mass distribution in the latitudinal direction, according to Fig. 3. $t_{\rm{c}}$ represents the time when $\theta_{\rm{obs}}$ becomes zero. At this time, $\int_{0}^{\theta_{\rm{ej}}} \rho \cos \theta d\theta =\frac{dM}{dv}/(2 \varphi_{\rm{ej}}\cdot r^{2}\cdot t)$, so $t_{\rm{c}}$ is independent of the mass distribution in the latitudinal direction. However, $t_{\rm{c}}$ should be dependent of the mass distribution with velocities. Below  we  calculate $t_{\rm{c}}$ by assuming $k=0$. The time when all of the ejecta becomes transparent is just the maximal $t_{\rm{c}}$ of blobs with different velocities.


The depth of the visible mass evolves with time as
\begin{equation}
\theta_{\rm{obs}}=\theta_{\rm{ej}}-\left[\frac{c}{\kappa \rho v^{2} t^{3}}\right]^{1/2} t=\theta_{\rm{ej}}\left(1-\frac{t}{t_{\rm{c}}}\right)
\end{equation}
where $t_{\rm{c}}=\theta_{\rm{ej}} [\frac{\kappa \rho v^{2} t^{3}}{c}]^{1/2}$.
For $s=0$, we have
\begin{equation}
\rho=\frac{M_{\rm{ej}}}{2\varphi_{\rm{ej}} \theta_{\rm{ej}} (v_{\rm{max}}-v_{\rm{min}})} v^{-2} t^{-3},
\end{equation}
so
\begin{equation}
t_{\rm{c}}=\left[\frac{\theta_{\rm{ej}} \kappa M_{\rm{ej}}}{2\varphi_{\rm{ej}}(v_{\rm{max}}-v_{\rm{min}})c}\right]^{1/2}.
\end{equation}

For $s=-1$, we have
\begin{equation}
\rho=\frac{M_{\rm{ej}}}{2\varphi_{\rm{ej}} \theta_{\rm{ej}} (\ln{v_{\rm{max}}}-\ln{v_{\rm{min}}})} v^{-3} t^{-3}
\end{equation}
and
\begin{equation}
t_{\rm{c}}=\left[\frac{\theta_{\rm{ej}} \kappa M_{\rm{ej}}}{2\varphi_{\rm{ej}} (\ln{v_{\rm{max}}}-\ln{v_{\rm{min}}})c
v_{\rm{min}}}\right]^{1/2}.
\end{equation}

For the reference values used in our model, we obtain $t_{\rm{c}}=4.87$d for $s=0$ and $t_{\rm{c}}=7.53$d for $s=-1$.
At $t>t_{\rm{c}}$, the visible masses for different cases approach to the same value, as shown in Figure 3. Correspondingly, the light curves of the bolometric luminosity approach to the same line, with $L\propto t^{-1.3}$.

\begin{figure}
\begin{center}
\includegraphics[scale=0.4]{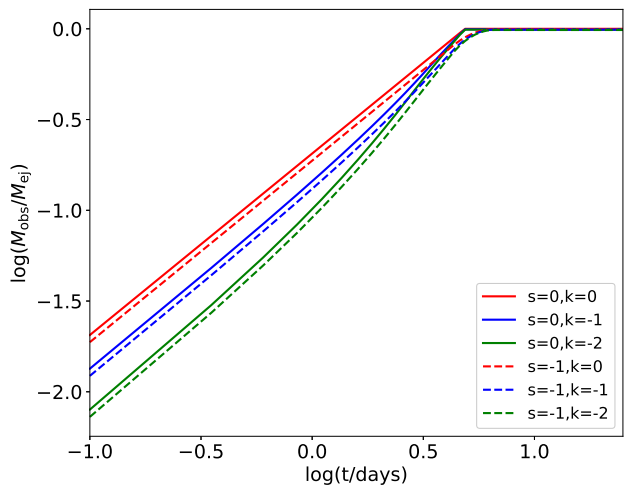}
\caption{The growth of  the visible mass $M_{\rm{obs}}$ with time. }
\end{center}
\end{figure}

The ejecta in our model can be regarded as the combination of blobs with different temperatures, which is different from  previous treatments that assume a single  temperature for the whole ejecta. We calculate the i-band light curves for various parameters shown in Fig. 4 to compare these two methods. The orange line is the the light curve when we ignore the anisotropic distribution in the latitudinal direction (i.e., $s=0$,  $k=0$) and assume  a single temperature for the ejecta,
\begin{equation}
T_{\rm{sin}} = \left(\frac{L}{\sigma_{\rm{SB}}S}\right)^{1/4},
\end{equation}
where $L$ is the total luminosity and $S=\varphi_{\rm{ej}}(R_{\rm{max}}^{2}-R_{\rm{min}}^{2})$ is the total surface area with $R_{\rm{max}}=v_{\rm{max}}t$, $R_{\rm{min}}=v_{\rm{min}}t$.
Comparing the light curve assuming blobs with different temperatures (the red solid line) and the case assuming a single temperature (the orange line) in Fig.4, we find the decay of the light curve is flatter in our model. The slopes of the dashed lines, the solid lines and the orange line at late time are $-0.70$, $-0.80$ and $-0.86$  respectively. Such a difference in the decay slopes could be discerned in the future with intensive observations.
The evolution at earlier times is mainly affected by the mass distribution in the latitudinal direction ($k$), because $k$ determine the increase of $M_{\rm{obs}}$ and thus determine the temperature of each blobs. After all of the ejecta becomes transparent, the mass distribution with the velocity ($s$) becomes important, and the lines with the same $s$ merge.

\begin{figure}
\begin{center}
\includegraphics[scale=0.4]{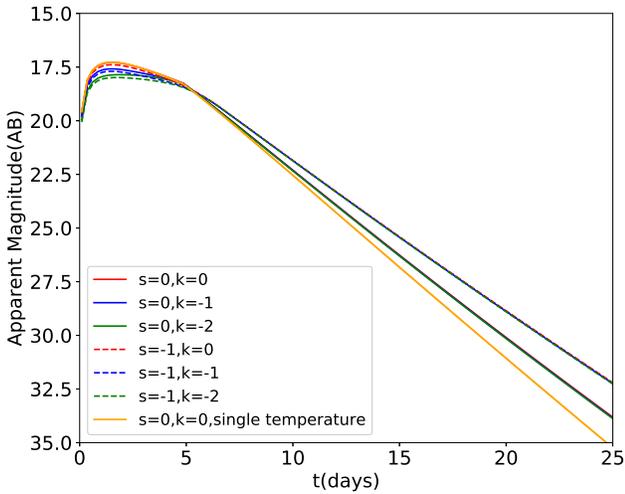}
\caption{The i-band light curves of the kilonova, where $D_{\rm{L}}=40$Mpc is adopted. The orange line is the magnitude evolution with time if we consider the ejecta as a blackbody with a single temperature $T_{\rm{sin}} = (\frac{L}{\sigma_{\rm{SB}}S})^{1/4}$, where $L$ is the total luminosity and $S$ is the total surface area. }
\end{center}
\end{figure}

{{In the above calculations, $v_{\rm{min}}=0.13c$ has been used for the $s=-1$ case. Note that $v_{\rm{min}}$ may affect the radiation from the kilonova. To check the importance of the minimal velocity as an input parameter, we take $v_{\rm{min}}=0.05c$ as an example, and compare it with the case of $v_{\rm{min}}=0.13c$  (for s=-1). The results are shown in Fig. 5. It can be seen that the values of $v_{\rm{min}}$ affect the bolometric luminosity of the kilonova at early times.  For a larger $v_{\rm{min}}$, the ejecta is denser. Therefore, it takes a longer time for the ejecta to become transparent and consequently the bolometric luminosity at early time is smaller. The i-band evolutions are also influenced by $v_{\rm{min}}$. The slopes at late time are -0.70 for $v_{\rm{min}}=0.13c$, and -0.52 for $v_{\rm{min}}=0.05c$,
respectively, as shown by the dashed line and the solid lines in Fig. 5 respectively.
\begin{figure*}
\begin{center}
\includegraphics[scale=0.39]{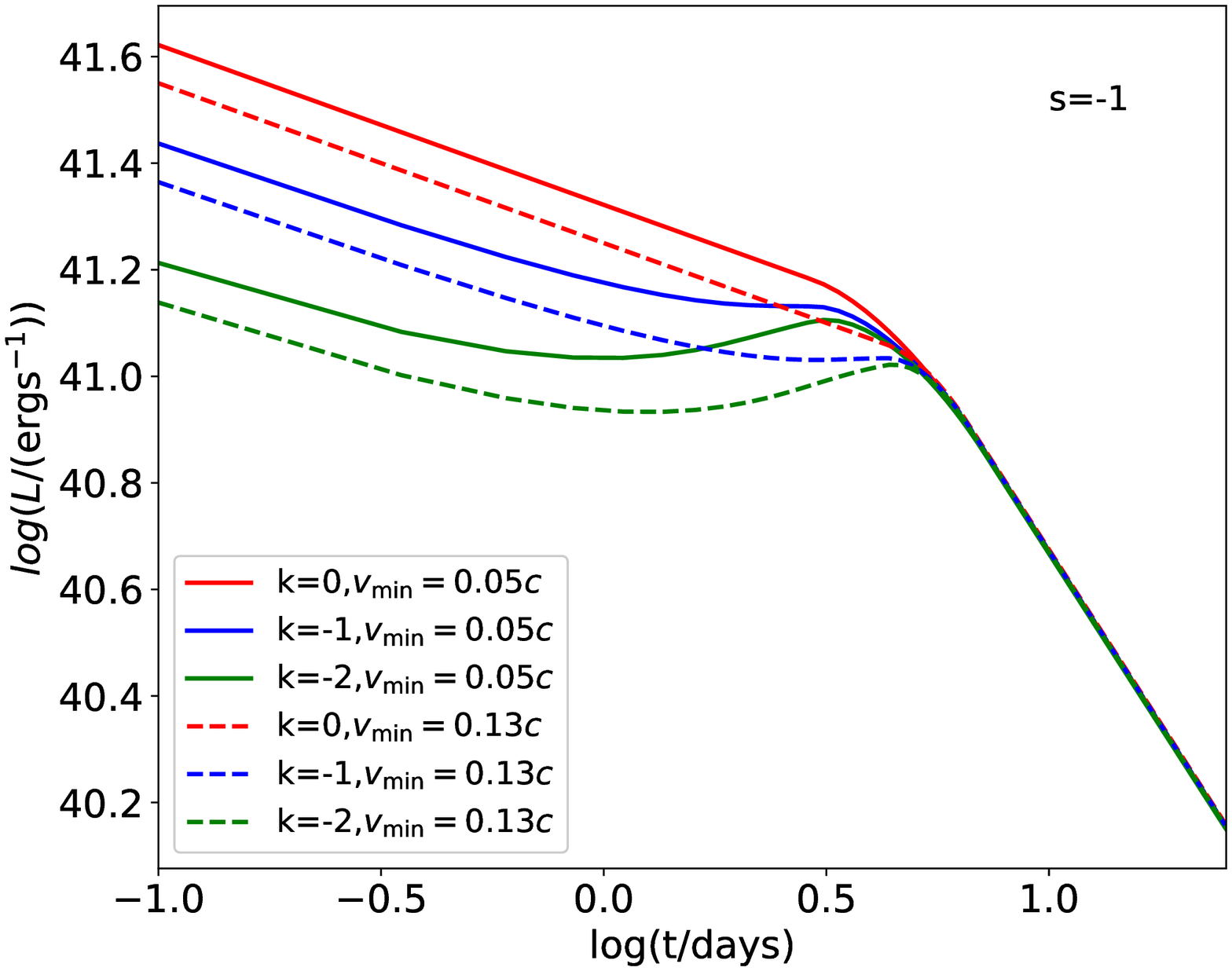}
\includegraphics[scale=0.39]{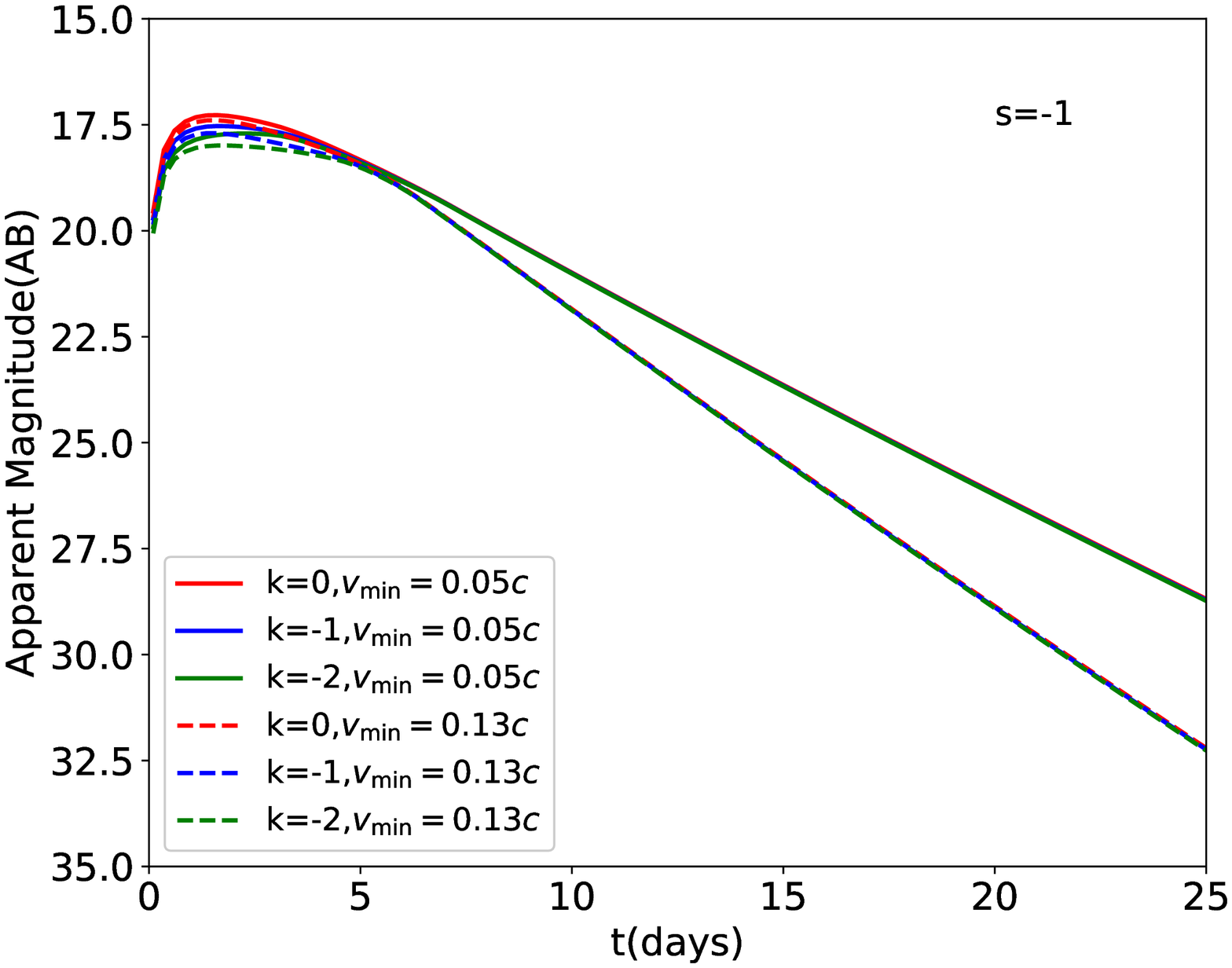}
\caption{Comparison between results for two different $v_{\rm{min}}$, given other parameters are the same. $s=-1$ is adopted. Left panel: Light curves of the bolometric luminosity of the kilonova. Right panel: The i-band light curves of the kilonova.}
\end{center}
\end{figure*}
}}

The radiation from the kilonova may depend on the viewing angle due to the anisotropic geometry\citep[e.g.,][]{2017CQGra..34o4001F}.
We  take the center of the ejecta as the origin and establish a spherical coordinate. For an off-axis observing angle $\theta _{\rm{ob}}$, a blob at $(r,\theta , \varphi)$ makes an angle $\alpha$ with respect to the observer. $\alpha$ is given by \citep{2018MNRAS.473L.121K}
\begin{equation}
\cos \alpha= \cos \theta _{\rm{ob}} \cos \theta + \sin \theta _{\rm{ob}} \sin \theta \cos \varphi.
\end{equation}
$\theta =0$ is the direction of the rotation axis. Since the ejecta has a small opening angle in the latitudinal direction, $\theta \sim \frac{\pi}{2}$ is adopted.

According to Eq. 5 in \citet{2017ApJ...851L..45G}, the observed flux contributed by a blob(Eq. 13) should be adapted as
\begin{equation}
\frac{dF_{\nu}^{'}}{d\varphi dv}= \frac{8\pi h \nu^{3}}{c^{2}} \frac{1}{exp(\frac{h\nu}{ {\mathcal{D}} k_{\rm{B}}T_{\rm{eff}}})-1} \frac{vt^{2}}{4\pi D_{L}^{2}} \frac{1}{{\mathcal{D}}^2},
\end{equation}
if the Doppler effect is taken into account, where ${\mathcal{D}}= 1/[\Gamma (1-\beta \cos \alpha)]$ is the Doppler factor.

The total flux contributed by the ejecta is
\begin{equation}
F_{\nu}^{'}= \int  \int _{v_{\rm{min}}}^{v_{\rm{max}}} \frac{dF_{\nu}^{'}}{d\varphi dv} d\varphi dv.
\end{equation}

Due to the crescent-like shape of the ejecta, the light curves vary with different directions of motion. Averaging over all directions of motion, half of the ejecta is assumed to move towards the observer while the others move away($\varphi \in [- \pi,0]$). Fig. 6 shows the i-band light curves with different viewing angles. The viewing angle can be independently obtained from the waveforms of the GWs \citep{2017PhRvL.119p1101A,2018ApJ...860L...2F} and their electromagnetic counterparts, like prompt emission \citep{2017ApJ...850L..24G} and afterglows \citep{2018PTEP.2018d3E02I,2018arXiv180609693M}. After knowing the viewing angle, the mass distribution in velocity can be further inferred from the light curves.

\begin{figure}
\begin{center}
\includegraphics[scale=0.41]{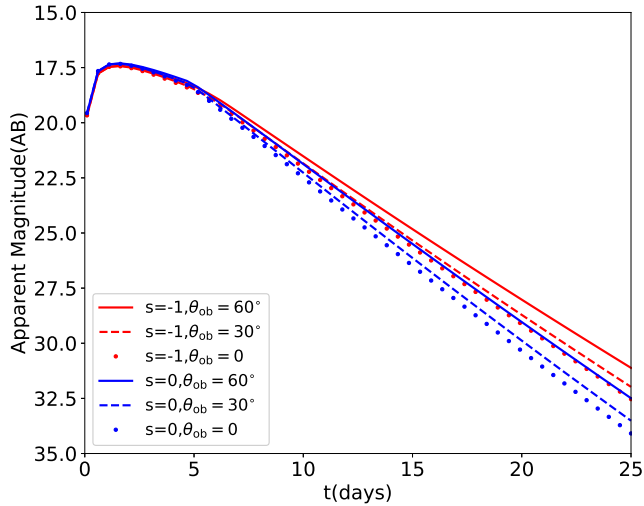}
\caption{The i-band light curves of the kilonova taking into account Doppler shift effects. Here $k=0$ is adopted. For each color, three viewing angles are shown, spanning the range from $\theta _{\rm{ob}}= 0$(rotation axis) to $\theta _{\rm{ob}}= 60^{\circ}$.}
\end{center}
\end{figure}

We also calculate the spectra before and after $t_{\rm{c}}$, as shown in Fig. 7. The shape of the spectra at earlier times (left panel) is mainly determined by $k$ since $k$ affects the growth rate of $M_{\rm{obs}}$.  $M_{\rm{obs}}$ and the bolometric luminosity are larger for $k=0$, thus the spectrum is wider. We can also see that, at earlier times, the differences between the spectrum considering the ejecta to have a single temperature (the orange solid line) and the case considering the ejecta to be a combination of blobs with different temperatures (the red solid line ) are small.
\begin{figure*}
\begin{center}
\includegraphics[scale=0.36]{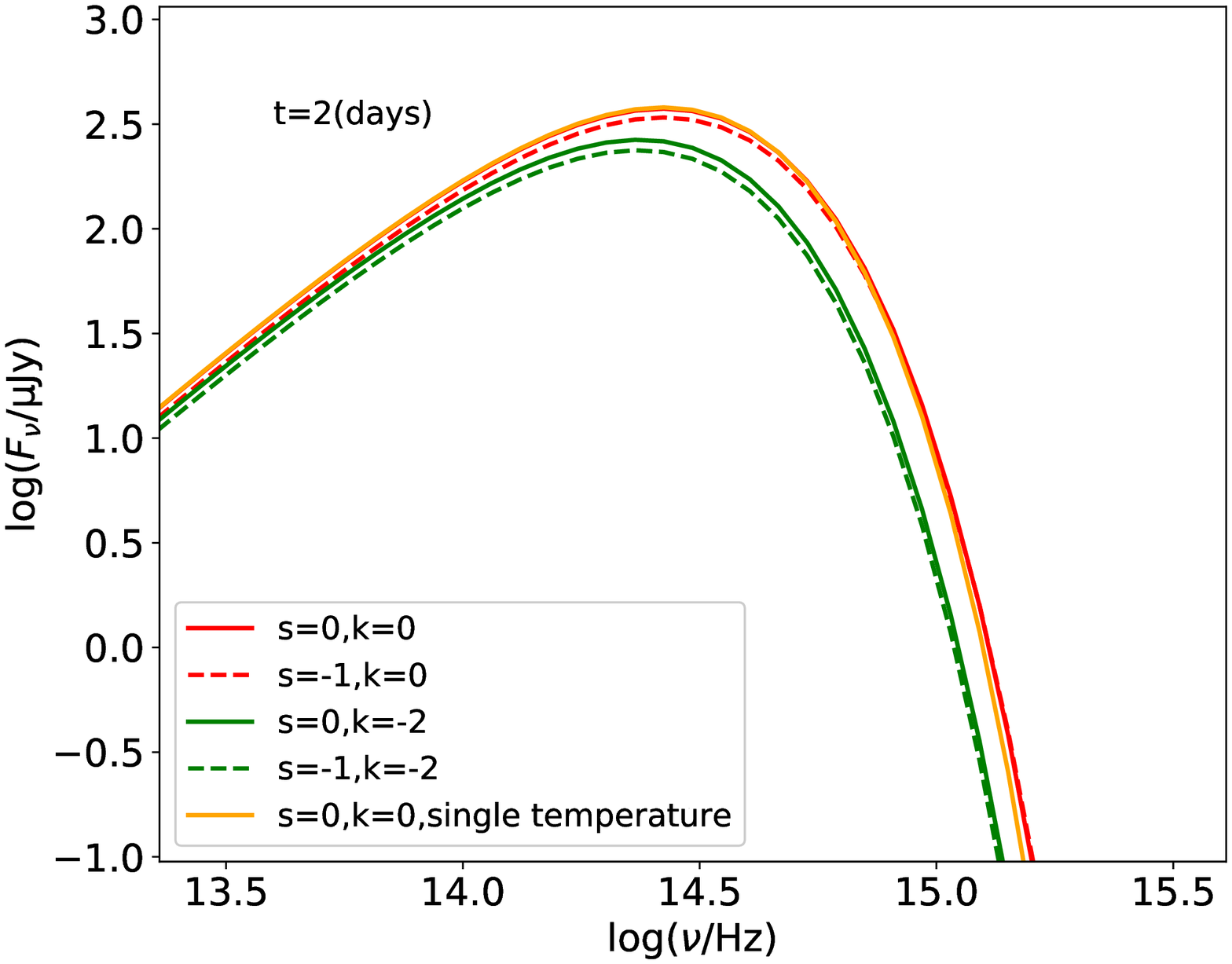}
\includegraphics[scale=0.36]{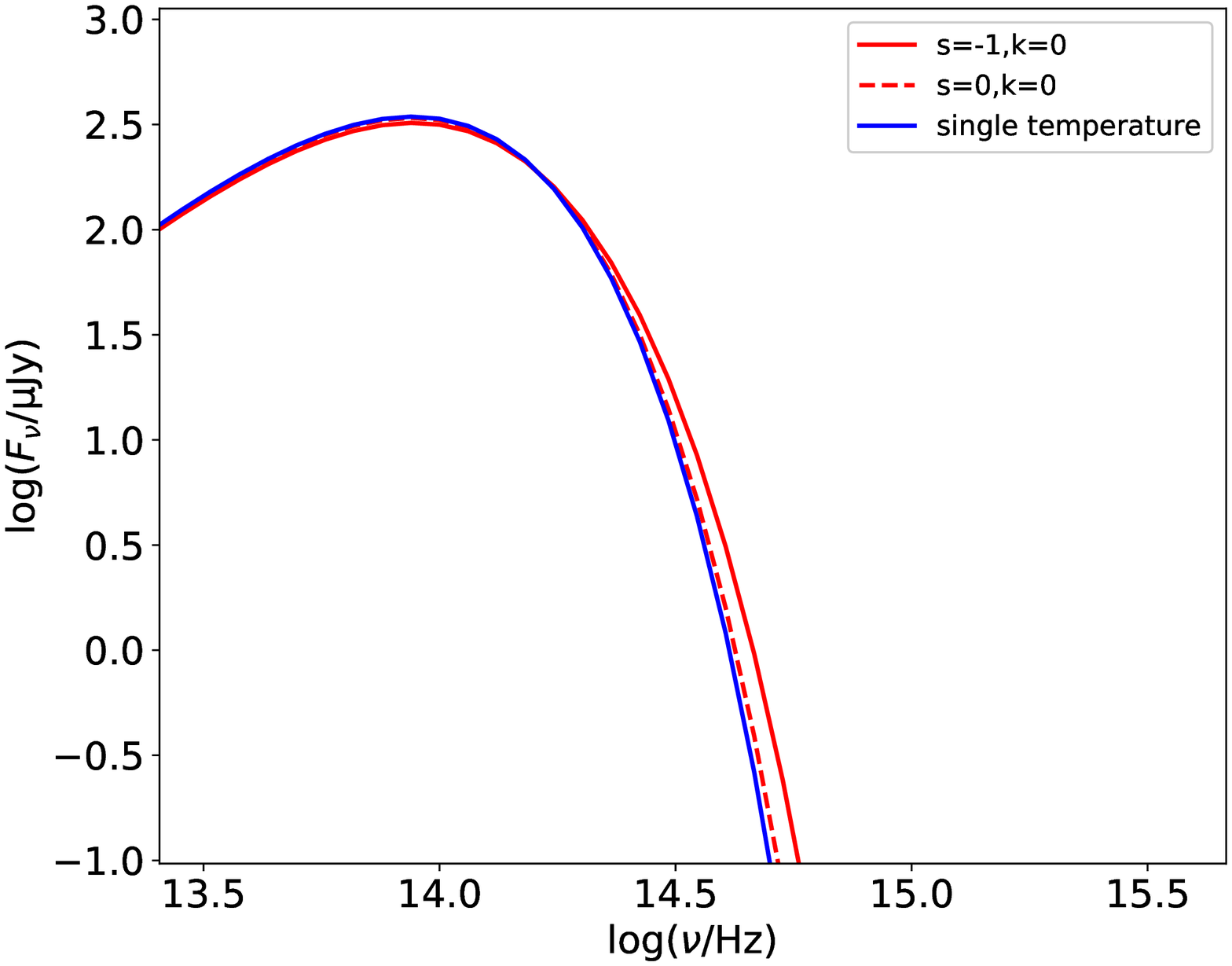}
\caption{Comparison between the spectra of the kilonova in our model and the case assuming a single temperature for the whole ejecta. Left panel: The spectra of the kilonova at $t=2$ day. The orange line assumes that the whole ejecta has  a single temperature of $T_{\rm{sin}} = (\frac{L}{\sigma_{\rm{SB}}S})^{1/4}$. Right panel: The spectra of the kilonova at $t=10$ day. The red solid line and the red dashed line represent the spectra calculated using our model for $s=-1$ and $s=0$, respectively. The blue line assumes that the whole ejecta has a single temperature.}
\end{center}
\end{figure*}

We also compare the spectra at $t=10$ day (the right panel), when all of the $M_{\rm{ej}}$ becomes transparent and the influence of $k$ can be neglected. The spectrum in our model is much harder at higher frequencies and seems difficult to be fitted by a blackbody spectrum, especially  for $s=-1$. This is because  more mass concentrates on lower velocity blobs and temperatures of these blobs are higher. Since these blobs mainly contribute to higher frequencies, the spectrum extends to higher frequencies. The temperature changes more significantly with velocity in the case of $s=-1$, so this effect is more obvious.

%
%
%
%

\section{Discussions and Conclusions}
Numerical simulations show that the mass distribution of the ejecta in  BH-NS mergers is highly anisotropic. There are also indications that the mass distribution  with velocity follows a power-law function. In this work, we study observational consequences of the kilonova emission due to this geometry anisotropy and the velocity distribution. Both of these factors can greatly influence the radiation from the kilonova. The kilonova emission at earlier times is sensitive to the mass distribution in the latitudinal direction (i.e., the $k$ value), since $k$ affects the growth of the visible mass $M_{\rm{obs}}$ with time before the whole ejecta becomes transparent. After the whole ejecta becomes transparent, the mass distribution in velocity (i.e., the $s$ value) heavily affects the emission spectrum and the light curve evolution. Due to the inhomogeneous mass distribution in the latitudinal direction, the increase of $M_{\rm{obs}}$ can be greater than the decay of the radioactive heating rate from  r-process nuclei, which causes a bump on the light curve of the bolometric luminosity. The existence of the bump can be an important feature to distinguish between BH-NS mergers and BNS mergers. Future detection of kilonova emission from BH-NS merger events would also be able to provide an opportunity to study the mass distribution profile of the merger ejecta.

Rather than considering the ejecta to have a single temperature, we treat the ejecta as a combination of blobs with different temperatures. This treatment causes differences in the i-band light curves and the emission spectra. We find that the i-band light curves becomes flatter and the spectrum is harder at high frequencies, which can be tested in future observations.

In our calculation, the opacity $\kappa$ is assumed to be  a constant. However, $\kappa$ may change with time and locations on the ejecta due to the evolution of the chemical composition and temperatures. This uncertainty will affect the growth of $M_{\rm{ej}}$ at early times. The opacity of the ejecta is dominated by the lanthanides and depend mildly on the lanthanide abundance \citep{2013ApJ...774...25K}. According to Fig. 10 in \cite{2013ApJ...774...25K}, the opacity can be treated as a constant when the temperature is higher than 4000K \citep{2017LRR....20....3M}, which is around $t=2-2.5$ day in our case. Later, the opacity decreases rapidly as the temperature drops. Therefore, the growth of $M_{\rm{obs}}$ is even faster than the case assuming $\kappa$ to be a constant. As a result, the bump on the bolometric luminosity light curve could be more significant. But the effect of the opacity on the i-band light curve and the emission spectrum at later times (i.e., when the whole ejecta becomes transparent) can be negligible. In future observations, the ejecta mass and the mass distribution in velocity can be derived by the i-band light curve at late times, while the bolometric luminosity at  early times can be used to determine the mass distribution in the latitudinal direction. After obtaining these parameters, one can break the degeneracy between the anisotropic mass ejection and the opacity effects  through numerical fittings with the observational data.

\acknowledgements We thank Kunihito Ioka, Koutarou Kyutoku and Masaomi Tanaka for valuable suggestions.
This work was supported by the National Basic Research Program of China (``973'' Program, Grant
No. 2014CB845800), the National Key Research and Development
Program of China (grant No.  2017YFA0402600 and 2018YFA0404200), and the National Natural Science Foundation of China (grant No.
11625312, 11851304and 11573014). L.D.L was supported by China Scholarship Program to conduct research at UNLV.

\end{document}